# Tin-Vacancy Quantum Emitters in Diamond

Takayuki Iwasaki,[1,*] Yoshiyuki Miyamoto,[2] Takashi Taniguchi,[3] Petr Siyushev,[4] Mathias H. Metsch,[4] Fedor Jelezko,[4,5] Mutsuko Hatano[1]

[1]Department of Electrical and Electronic Engineering, Tokyo Institute of Technology, Meguro, Tokyo 152-8552, Japan

[2]Research Center for Computational Design of Advanced Functional Materials, National Institute for Advanced Industrial Science and Technology

[3]Advanced Materials Laboratory, National Institute for Material Science, 1-1 Namiki, Tsukuba, 305-0044, Japan

[4]Institute for Quantum Optics, Ulm University, D-89081 Germany

[5]Center for Integrated Quantum Science and Technology (IQ[st]), Ulm University, D-89081 Germany

e-mail: iwasaki.t.aj@m.titech.ac.jp

**Abstract**

Tin-vacancy (SnV) color centers were created in diamond by ion implantation and subsequent high temperature annealing up to 2100 °C at 7.7 GPa. The first-principles calculation suggests that the large atom of tin can be incorporated into the diamond lattice with a split-vacancy configuration, in which a tin atom sits on an interstitial site with two neighboring vacancies. The SnV center shows a sharp zero phonon line at 619 nm at room temperature. This line splits into four peaks at cryogenic temperatures with a larger ground state splitting of ~850 GHz than those of color centers based on other IV group elements, silicon-vacancy (SiV) and germanium vacancy (GeV) centers. The excited state lifetime was estimated to be ~5 ns by Hanbury Brown-Twiss interferometry measurements on single SnV quantum emitters. The order of the experimentally obtained optical transition energies comparing with the SiV and GeV centers is good agreement with the theoretical calculations.

Point defect-related color centers in solid state materials are a promising approach for quantum information processing [1,2]. Nitrogen-vacancy (NV) centers in diamond have been most intensively studied from the viewpoint of both fundamental and applied sciences [3,4]. However, the zero phonon line (ZPL) of the NV center possesses a fraction of only 4 % in its total fluorescence due to its large phonon sideband (PSB). Also, the NV center suffers from external noise, leading to instability of the optical transition energy. To overcome these drawbacks, color centers based on IV group elements, silicon-vacancy (SiV) [5-7] and germanium-vacancy (GeV) [8-11] centers, has attracted attention owing to their large ZPL, structural symmetry robust against the external noise, and availability of quantum emission by single centers. Recently, spin control and evaluation of spin coherence time have been investigated for this two color centers [12-16], revealing that their spin coherence times were limited to sub-microseconds even at cryogenic temperatures < 5 K, much shorter than that of the NV center [17,18]. This limitation originates from phonon-mediated transitions between the lower and upper branches in the ground state [13-16]. Further cooling down to the sub-Kelvin regime or strain engineering is considered as a solution [15,16]. Another possibility is creation of a novel color center possessing a larger energy splitting in the ground state. For this purpose, in this study, we investigated the utilization of a IV group atom of tin (Sn, Fig. 1a). The incorporation of such a heavy atom into the diamond lattice as a form of a color center, especially single quantum emitter, has not been demonstrated yet. Here, we fabricated tin-vacancy (SnV) centers in diamond in both ensemble and single states by combination of ion implantation and subsequent high temperature annealing up to 2100 °C under a high pressure of 7.7 GPa. Low temperature optical measurements revealed that the ground state splitting of the SnV center was much larger than those of the SiV and GeV centers. The atomic structure and optical transition energy were calculated by the first-principle calculations with comparing to the SiV and GeV centers.

We used IIa-type single-crystal (001) diamond substrates with a low nitrogen concentration < 5 ppb (electronic grade supplied by element six). The Sn ions were implanted into diamond with doses of $2 \times 10^8 – 2 \times 10^{13}$ cm$^{-2}$ at acceleration energies of 130 – 150 keV, giving rise to a projected depth of approximately 40 nm from the diamond surface. Then, the samples were annealed at high temperatures ranging from 800 to 2100 °C under either a high pressure of 7.7 GPa or a high vacuum of ~$10^{-5}$ Pa. The high pressure was applied in pressure transmitting medium of CeCl by using belt-type high pressure apparatus. The vacuum annealing was done in an infrared heating furnace. Optical properties were evaluated by micro-Raman (inVia confocal Raman microscope, Renishaw) and home-built confocal fluorescence microscope systems at room temperature and low temperatures down to 4 K. The excitation laser wavelength was 532 nm. Hanbury Brown-Twiss (HBT) interferometry [19] measurements were performed using two avalanche photo diodes.

The atomic structure, energy levels, and optical transition energy were estimated by first-principles calculations. We used cubic 4x4x4 cell with equivalent carbon site number is 512. Planewave basis set with cutoff energy of 60 Ry, and norm-conserving pseudopotentials [20] were used. For the exchange-correlation potential, the PBE functional [21] was employed. The momentum space integration was performed using the Γ point, and the geometry optimization was performed with force criterion less than 0.05 eV/Å.

Figure 1b shows a room temperature PL spectrum from ensemble SnV centers, annealed at 2100 °C at 7.7 GPa after Sn ion implantation into diamond. The SnV centers show a sharp and strong ZPL at 619 nm with a full width at half maximum (FWHM) of 6.2 nm, accompanying with PSB. We investigated the effect of the post-implantation annealing temperature on the activation of the SnV centers. Figure 1c shows temperature dependence of the linewidth of ZPL in the SnV spectrum. We have performed the annealing under vacuum or high pressure conditions. The treatment duration was 30 min for all samples. The increase in the annealing temperature effectively reduces the FWHM irrespective of the annealing environments. Significantly, the high pressure of 7.7 GPa enabled us to reach a temperature of 2100 °C with avoiding the graphitization of the diamond [22,23]. Consequently, the linewidth was reduced from ~12 nm at 800 °C down to ~6 nm at 2100 °C by suppressing inhomogeneous broadening due to surrounding defects created during ion implantation and strain in the atomic structure. Furthermore, such a high temperature treatment is important to suppress the formation of undesirable fluorescent structures. Although two prominent peaks with unknown origins were observed at 595 and 646 nm at temperatures of 1500 °C and lower, they were annealed out by the treatment at 2100 °C (Supplemental Material). The large Sn atom is thought to be less movable in the diamond lattice and to produce a number of vacancies during ion implantation. Thus, the high temperature treatment is required to selectively activate the high-quality SnV centers.

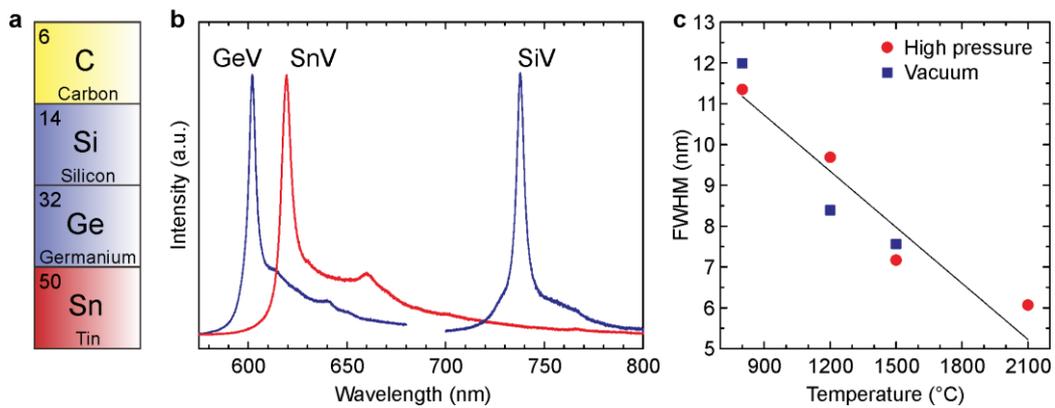

Figure 1. Activation and optical properties of ensemble SnV centers. (a) The IV group periodic table. (b) Room temperature PL spectrum from ensemble of SnV centers, annealed at 2100 °C after ion implantation with a dose of $2 \times 10^{13}$ cm$^{-2}$ at an acceleration energy of 130 keV. PL spectra from ensemble of SiV and GeV centers are also shown. (c) Dependence of linewidth of ZPL from SnV on the post-implantation annealing temperature under a high vacuum of ~$10^{-5}$ Pa or high pressure of 7.7 GPa. The annealing time was 30 min.

Low temperature PL measurements were performed to reveal the fine structure of the SnV center (Fig. 2a). Only the one peak is observed at room temperature, while it splits into two peaks with FWHM of approximately 0.2 nm at 4 K. To investigate the fine structure in details, the temperature dependence of the PL spectrum was recorded, as shown in Fig. 2b. When increasing the temperature, another two peaks (marked as C and D) appeared in a higher energy region in addition to the peaks seen at 4 K (marked as A and B). The intensities of the C and D peaks become larger as increasing the temperature. This fact suggests that the SnV center has an energy level structure composed of four levels with split ground- and excited-states, as shown in Fig. 2c. The elevated temperatures lead to thermal excitation in the excited state [5], and thus, the intensities of the C and D peaks become stronger with respect to the A and C peaks, respectively. The SiV and GeV centers also possess the same four level structure [5, 9], but with different splittings. In Fig. 2d, we summarize the splittings as a function of the atomic number of the IV group elements. Both the ground and excited state splittings increase as the atomic-size of the elements becomes larger. The ground and excited state splittings of the SnV center are ~850 and ~3000 GHz, respectively. This ground state splitting of the SnV center is 17 and 5 times larger than those of SiV and GeV, respectively. The large splittings obtained here come from the stronger spin-orbit interaction in a heavier atom. In addition, the dynamic Jahn-Teller effect would also contribute to the splittings, as demonstrated in the SiV centers [24]. The degree of the two effects will be revealed by observing the fine structures under magnetic fields [24].

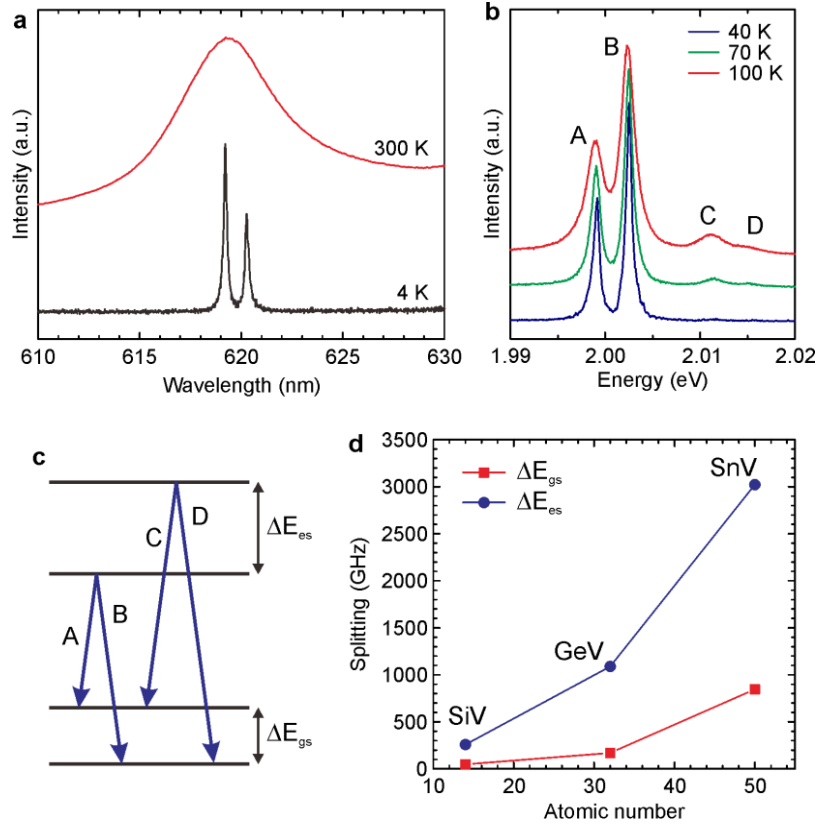

Figure 2. Low temperature optical characteristics. (a) ZPL of ensemble SnVs at room temperature and at 4 K. (b) PL spectra at different measurement temperatures. (c) Fine structure. (d) Ground and excited state splittings as a function of the atomic number of the IV group element. The values of the SiV and GeV centers were adapted from Refs. [5] and [9], respectively.

We next fabricated single SnV color centers. The inset in Fig. 3a shows a confocal fluorescence microscopy image of an isolated SnV center with FWHM of ~450 nm. PL spectrum was recorded at the bright spot and background position in Fig. 3a. The bright spot clearly shows a ZPL with a linewidth of 6 nm from the SnV center. The background position also shows multiple peaks at the same region, but they originate from the second-order Raman from diamond (611 and 620 nm) and probably a surface defect (630 nm) which is not seen in the bulk diamond region. The background-subtracted PL shows ZPL and PSB from only the SnV center. We performed the HBT interferometry measurements to confirm that the observed spots correspond to a single SnV center. Figure 3b shows background corrected [25] second-order autocorrelation function, $g^2(\tau)$, at different excitation laser powers. The $g^2(\tau)$ almost reaches zero at a delay time of 0 ns, which is an unambiguous proof of the single photon emission [26] from the observed SnV center. The data were fitted with an equation [27], $g^2(\tau) = 1 - (1+\alpha)e^{-\frac{|\tau|}{\tau_1}} + \alpha e^{-\frac{|\tau|}{\tau_2}}$, where α, $\tau_1$, and $\tau_2$ are fitting parameters. $\tau_1$ gives an estimated excited state lifetime of a fluorescent structure. Two emitters investigated in this study show $\tau_1$ of ~5 ns (another emitter is shown in Supplemental Material). At all the laser powers used here, $g^2(\tau)$ shows bunching behavior seen as $g^2(\tau)$ over the unity. This means that the SnV center consists of a three-level system in addition to the ground- and excited-states or that photoionization occurs during the laser excitation [27].

The fluorescence intensity was measured as a function of the laser power (Fig. 3c). We used two kinds of filters: Only the ZPL was monitored with a band-pass filter (BPF) with 20 nm width around the ZPL, while the intensity from the whole SnV spectrum (ZPL+PSB) was recorded with a 600 nm long-pass filter (LPF). Both curves show non-linear behavior at high laser powers. The plots were fitted with an equation [6], $I = I_\infty \times P/(P + P_{sat})$, where $I_\infty$ and $P_{sat}$ are saturation intensity and saturation power, respectively. The saturation intensities of the ZPL and whole spectrum are 280 kcps and 530 kcps, respectively. Note that we used an air-objective here, so the utilization of an oil-objective and/or microcavity structures [28-31] would enhance the light collection efficiency from the emitter and lead to the further increase in the saturation intensity. It is worth noting that the comparison of the fluorescence intensity of the ensemble with the single SnV center fabricated by the 2100 °C annealing provides a conversion efficiency from Sn ions to SnV centers of approximately 1-2 %.

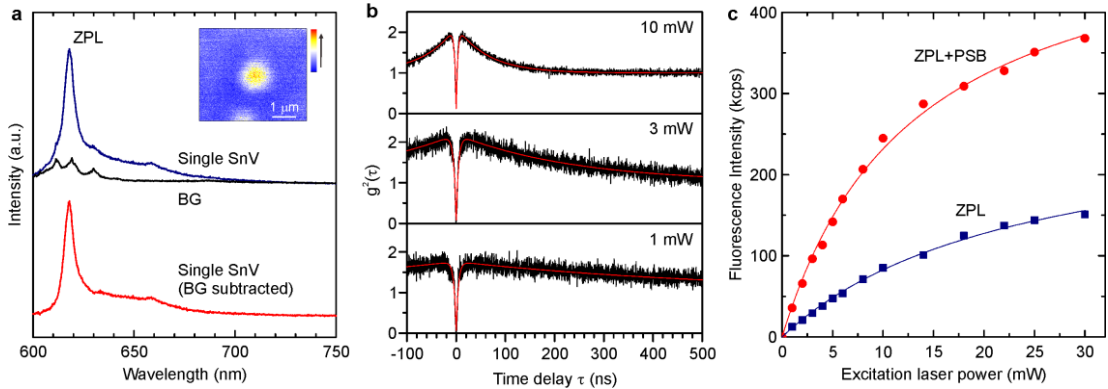

Figure 3. Single SnV color center. (a) PL spectrum from an isolated SnV center shown in inset. Inset: confocal fluorescence microscopy image taken with a BPF with 20 nm width around the ZPL. BG denotes PL at the background position where no SnV center exists. The bottom curve is background-subtracted single SnV spectrum. (b) Background corrected second-order autocorrelation function, $g^2(\tau)$, at different excitation laser powers, measured with the BPF. (c) Saturation curves of only ZPL and whole spectrum. The background intensities were subtracted from each curve. The single SnV centers were fabricated by ion implantation (dose: $2 \times 10^8$ cm$^{-2}$, acceleration energy: 150 keV) and annealing at 2100 °C under a high pressure of 7.7 GPa. All the measurements were done with an air-objective at room temperature.

We have performed the first-principles calculations of the atomic structure, energy levels, and optical transition energy of the SnV center. Figure 4a shows the optimized atomic configuration of the SnV center taking a split-vacancy structure. The Sn atom sits at an interstitial site between two vacancies at carbon lattice sites. This structure possesses $D_{3d}$ symmetry, same as the SiV and GeV centers. In accordance with the order of atomic sizes, the nearest C-Si, C-Ge, and C-Sn distances are 1.94, 2.05 and 2.14 Å, respectively. The energy levels of the ground state of the SnV center assuming in the negatively-charged state are shown in Fig. 4b. The degenerated $e_u$ levels are fully occupied while one of the degenerated $e_g$ levels are half-occupied. Upon the optical excitation, one electron with down spin in the $e_u$ level is pumped up to the $e_g$ level, and then relaxes back while emitting photons with an energy of ~2 eV.

Here, we discuss the optical transition energy of the color centers based on the IV group elements. Interestingly, irrespective of having the same structural symmetry, the order of the optical transition energy is not proportional to the atomic size of the element. Figure 1b also shows PL spectra from SiV and GeV centers in addition to the SnV center. The SiV centers possess ZPL at 738 nm, while the GeV centers composed of a larger atom show a higher transition energy at 602 nm. However, the ZPL energy of the SnV centers with an even heavier atom is less than that of the GeV center. The transition energy of the SnV was calculated using the structural model in Fig. 4a as well as SiV and GeV, which are summarized in Fig. 4c including the experimental ZPL transition energies. The experimental order of the optical excitation gap (SiV < SnV < GeV) is consistent with theoretical results, which was also reported by Goss et al. [32].

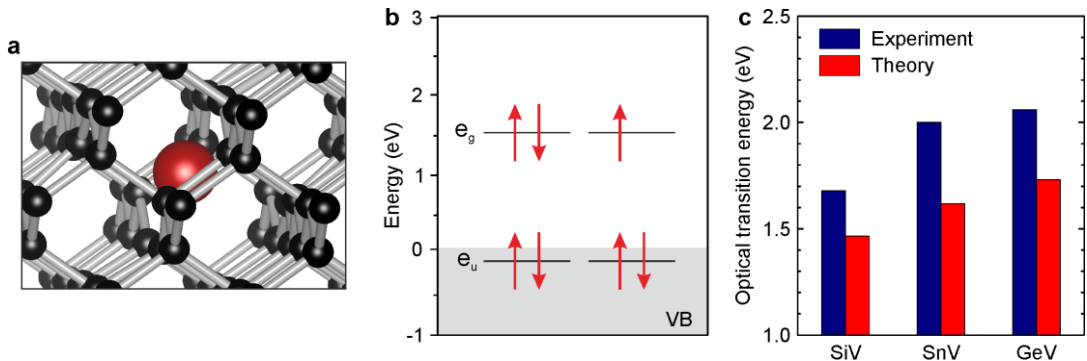

Figure 4. Theoretical calculation of SnV center by first-principles calculation. (a) Atomic structural model. A red sphere and black spheres denote a Sn atom and C atoms, respectively. (b) Energy levels of the ground state. VB denotes valence band of diamond. (c) Optical transition energies comparing with SiV and GeV centers.

We introduced large Sn atoms in the diamond lattice and obtained the SnV centers with a large level splitting in the ground state. Incorporation of a heavier atom than those reported in previous studies [33] in diamond will become more important from the viewpoints of material science and quantum optics. The incorporation of large atoms such as Eu [34] and Xe [35] has been demonstrated so far. For the further development of the heavy atom-related color centers, it is essential to achieve high-quality quantum emitters by the adequate process for each atom, such as high temperature annealing under a high pressure as demonstrated in this study.

In conclusion, we demonstrated the fabrication of both ensemble and single SnV centers in diamond by ion implantation and annealing. The high annealing temperature of 2100 °C under a high pressure of 7.7 GPa led to the narrow ZPLs from the SnV centers while annealing out undesirable fluorescent structures. Low temperature optical measurements revealed that the SnV center possess a four-level structure with a large splitting in the ground state, ~850 GHz. Comparing with the SiV and GeV centers, we showed that the magnitude of the splitting is in proportional with the atomic number of the IV group element owing to the stronger spin-orbit interaction. Our theoretical calculations indicated the split-vacancy configuration of the SnV center in agreement with the observed fine structures. The development of the color center with a large ground state splitting is expected to provide a way to establish the quantum light-mater interface with long spin coherence time.

Note: During the preparation of this manuscript, we became aware of another study showing room temperature characteristics of Sn-related color centers in diamond [36].

**Acknowledgements**
This work was supported by JST-PRESTO Grant Number JPMJPR16P2 and JST-CREST. We thank Shinji Nagamachi for supporting ion implantation.

**References**
[1]  R. Hanson and D. Awschalom, Nature 453, 1043 (2008).
[2]  I. Aharonovich, D. Englund, and M. Toth, Nat. Photonics 10, 631 (2016).
[3]  F. Jelezko and J. Wrachtrup, phys. stat. sol. (a), 203, 3207 (2006).
[4]  T. Schröder, S. L. Mouradian, J. Zheng, M. E. Trusheim, M. Walsh, E. H. Chen, L. Li, I. Bayn, and D. Englund, J. Opt. Soc. Am. B 33, B65 (2016).
[5]  C. D. Clark, H. Kanda, I. Kiflawi, and G. Sittas, Phys. Rev. B 51, 16681 (1995).
[6]  C. Wang, C. Kurtsiefar, H. Weinfurter, and B. Burchard, *J. Phys. B: At. Mol. Opt. Phys*. 39, 37 (2006).
[7]  A. Sipahigil, R. E. Evans, D. D. Sukachev, M. J. Burek, J. Borregaard, M. K. Bhaskar, C. T. Nguyen, J. L. Pacheco, H. A. Atikian, C. Meuwly, R. M. Camacho, F. Jelezko, E. Bielejec, H. Park, M. Lončar, and M. D. Lukin, Science 354, 847 (2016).

[8]  T. Iwasaki, F. Ishibashi, Y. Miyamoto, Y. Doi, S. Kobayashi, T. Miyazaki, K. Tahara, K. D. Jahnke, L. J. Rogers, B. Naydenov, F. Jelezko, S. Yamasaki, S. Nagamachi, T. Inubushi, N. Mizuochi, and M. Hatano, Sci. Rep. 5, 12882 (2015).

[9]  Y. N. Palyanov, I. N. Kupriyanov, Y. M. Borzdov, and N. V. Surovtsev, Sci. Rep. 5, 14789 (2015).

[10] E. A. Ekimov, S. G. Lyapin, K. N. Boldyrev, M. V. Kondrin, R. Khmelnitskiy, V. A. Gavva, T. V. Kotereva, and M. N. Popova, JETP Lett. 102, 701 (2015).

[11] M. K. Bhaskar, D. D. Sukachev, A. Sipahigil, R. E. Evans, M. J. Burek, C. T. Nguyen, L. J. Rogers, P. Siyushev, M. H. Metsch, H. Park, F. Jelezko, M. Lončar, and M. D. Lukin, Phys. Rev. Lett. 118, 223603 (2017).

[12] T. Müller, C. Hepp, B. Pingault, E. Neu, S. Gsell, M. Schreck, H. Sternschulte, D. Steinmüller-Nethl, C. Becher, and M. Atatüre, Nat. Comm. 5, 3328 (2014).

[13] B. Pingault, J. N. Becker, C. H. H. Schulte, C. Arend, C. Hepp, T. Godde, A. I. Tartakovskii, M. Markham, C. Becher, and M. Atatüre, Phys. Rev. Lett. 113, 263601 (2014).

[14] L. J. Rogers, K. D. Jahnke, M. H. Metsch, A. Sipahigil, J. M. Binder, T. Teraji, H. Sumiya, J. Isoya, M. D. Lukin, P. Hemmer, and F. Jelezko, Phys. Rev. Lett. 113, 263602 (2014).

[15] B. Pingault, D.-D. Jarausch, C. Hepp, L. Klintberg, J. N. Becker, M. Markham, C. Becher, and M. Atatüre, Nat. Comm. 8, 15579 (2017).

[16] P. Siyushev, M. H. Metsch, A. Ijaz, J. M. Binder, M. K. Bhaskar, D. D. Sukachev, A. Sipahigil, R. E. Evans, C. T. Nguyen, M. D. Lukin, P. R. Hemmer, Y. N. Palyanov, I. N. Kupriyanov, Y. M. Borzdov, L. J. Rogers, and F. Jelezko, arXiv:1612.02947.

[17] G. Balasubramanian, P. Neumann, D. Twitchen, M. Markham, R. Kolesov, N. Mizuochi, J. Isoya, J. Achard, J. Beck, J. Tissler, V. Jacques, P. R. Hemmer, F. Jelezko, and J. Wrachtrup, Nat. Mater. 8, 383 (2009).

[18] N. Bar-Gill, L.M. Pham, A. Jarmola, D. Budker, and R.L. Walsworth, Nat. Comm. 4, 1743 (2013).

[19] R. Hanbury and R. Q. Twiss, Nature 178, 1046 (1956).

[20] N. Troullier and J. L. Martins, Phys. Rev. B 43, 1993 (1991).

[21] J. P. Perdew, K. Burke, and M. Ernzerhof, Phys. Rev. Lett. 77, 3865 (1996).

[22] A. T. Collins, A. Connor, C. -H. Ly, and A. Shareef, J. Appl. Phys. 97, 083517 (2005).

[23] B. Harte, T. Taniguchi, and S. Chakraborty, Mineralogical Magazine 73, 201 (2009).

[24] C. Hepp, T. Müller, V. Waselowski, J. N. Becker, B. Pingault, H. Sternschulte, D. Steinmüller-Nethl, A. Gali, J. R. Maze, M. Atatüre, and C. Becher, Phys. Rev. Lett. 112, 036405 (2014).

[25] R. Brouri, A. Beveratos, J.-P. Poizat, and P. Grangier, Opt. Lett. 25, 1294 (2000).

[26] M. Leifgen, T. Schröder, F. Gädeke, R. Riemann, V. Métillon, E. Neu, C. Hepp, C. Arend, C. Becher, K. Lauritsen, and O. Benson. New J. Phys. 16, 023021 (2014).

[27] S. Prawer and I. Aharonovich, Quantum Information Processing with Diamond: Principles and Applications. ch. 6 (Woodhead Publishing, 2014).

[28] J. P. Hadden, J. P. Harrison, A. C. Stanley-Clarke, L. Marseglia, Y.-L. D. Ho, B. R. Patton, J. L. O'Brien, and J. G. Rarity, Appl. Phys. Lett. 97, 241901 (2010).

[29] M. Jamali, I. Gerhardt, M. Rezai, K. Frenner, H. Fedder, and J. Wrachtrup, Rev. Sci. Instrum. 85, 123703 (2014).

[30] T. M. Babinec, B. J. M. Hausmann, M. Khan, Y. Zhang, J. R. Maze, P. R. Hemmer, and M. Lončar, Nat. Nanotech. 5, 195 (2010).

[31] S. Furuyama, K. Tahara, T. Iwasaki, M. Shimizu, J. Yaita, M. Kondo, T. Kodera, and M. Hatano, Appl. Phys. Lett. 107, 163102 (2015).

[32] J. P. Goss, P. R. Briddon, M. J. Rayson, S. J. Sque, and R. Jones, Phys, Rev. B 72, 035214 (2005).

[33] V. Nadolinny, A. Komarovskikh, and Y. Palyanov, Crystals 7, 237 (2017).

[34] A. Magyar, W. Hu, T. Shanley, M. E. Flatte, E. Hu, and I. Aharonovich, Nat. Comm. 5, 3523 (2014).

[35] R. Sandstrom, L. Ke, A. Martin, Z. Wang, M. Kianinia, B. Green, W. −B. Gao, I. Aharonovich, arXiv:1704.01636.

[36] S. D. Tchernij, T. Herzig, J. Forneris, J. Küpper, S. Pezzagna, P. Traina, E. Moreva, I. P. Degiovanni, G. Brida, N. Skukan, M. Genovese, M. Jakšić, J. Meijer, and P. Olivero, arXiv:1708.01467.

# Supplemental Materials

# for Tin-Vacancy Quantum Emitters in Diamond

Takayuki Iwasaki,[1,*] Yoshiyuki Miyamoto,[2] Takashi Taniguchi,[3] Petr Siyushev,[4] Mathias H. Metsch,[4] Fedor Jelezko,[4,5] Mutsuko Hatano[1]

[1]Department of Electrical and Electronic Engineering, Tokyo Institute of Technology, Meguro, Tokyo 152-8552, Japan

[2]Research Center for Computational Design of Advanced Functional Materials, National Institute for Advanced Industrial Science and Technology

[3]Advanced Materials Laboratory, National Institute for Material Science, 1-1 Namiki, Tsukuba, 305-0044, Japan

[4]Institute for Quantum Optics, Ulm University, D-89081 Germany

[5]Center for Integrated Quantum Science and Technology (IQ$^{st}$), Ulm University, D-89081 Germany

**1. Dependence of PL spectra from the SnV centers on the annealing temperature**

Figure S1a,b show PL spectra from Sn ion implanted diamonds followed by annealing at different temperatures under high pressure and vacuum. For both the environments, in addition to the ZPL from the SnV center at 619 nm, other two strong peaks appear depending on the annealing temperature. One peak is located at 595 nm which is mainly observed at 800 °C. The 595 nm peak is completely invisible at 1500 °C and higher. Another peak at 646 nm becomes most intensive at 1500 °C and disappears at 2100 °C. Importantly, the high temperature of 2100 °C under high pressure only makes it possible to selectively obtain the fluorescence from the SnV center.

It is worth noting that the 594 nm "absorption" peak has been reported in nitrogen-containing type Ib diamonds after neutron or electron irradiation and annealing [1,2]. This peak is thought to be related with interstitial nitrogen atoms [1]. Furthermore, it was shown that this peak was inactive in luminescence [2]. Thus, the 595 nm peak observed in this study is not likely to be identical to the 594 nm peak observed in absorption. We show low temperature PL spectra of the 595 nm peak, which might become a hint to clarify the origin (Fig. S1d). We see two peaks at 10 K, while four peaks appear at a higher temperature of 60 K. This phenomenon is similar with the ZPL from the SnV center. Thus, a four-level structure with the ground and excited state splittings is one possible energy level. The ground state splitting is much lower than that of the SnV, but it is 1.6 times higher than that of the GeV. This structure might be related with Sn, for example forming an intermediate state before being incorporating into the perfect interstitial site.

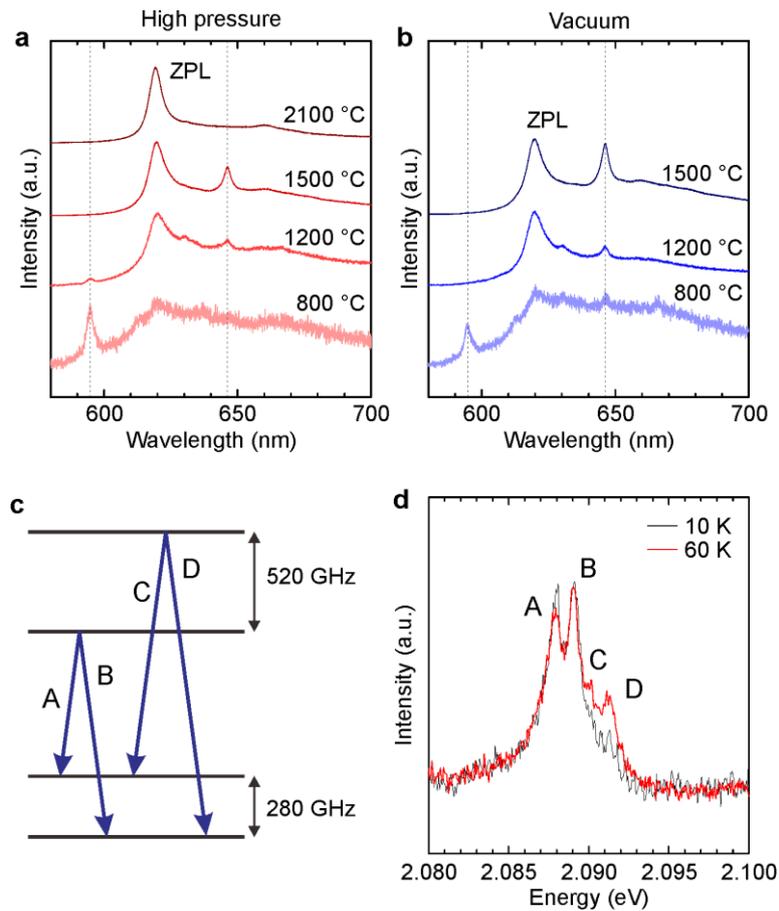

Figure S1. PL spectra from ensemble SnV centers annealed at different temperatures under (a) a high pressure of 7.7 GPa and (b) vacuum. (c) Possible energy levels and (d) low temperature PL spectra of the 595 nm center.

## 2. Another single SnV center

Figure S2 show another single SnV center. The $g^2(0)$ is close to zero at a delay of 0 ns, indicating the quantum emission from the SnV center.

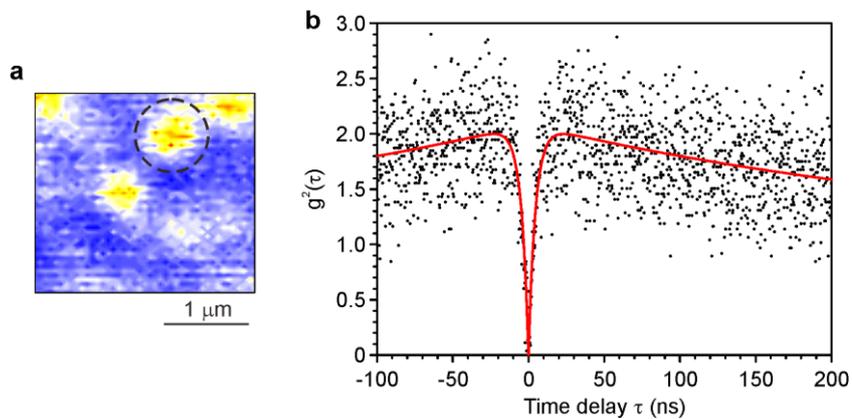

Figure S2. (a) Confocal fluorescence microscope image. (b) Background-corrected $g^2(\tau)$ from the fluorescent spot marked in circle in panel a. The measurements were performed at an excitation laser power of 3 mW with a 600 nm LPF.

**References**

[1] Y. Nishida, Y. Mita, K. Mori, S. Okuda, S. Sato, S. Yazu, M. Nakagwa, and M. Okada, Mater. Sci. Forum 38-41, 561 (1989).
[2] A. T. Collins, A. Connor, C. -H. Ly, and A. Shareef, J. Appl. Phys. 97, 083517 (2005).